\documentclass{article}%
\usepackage{amsfonts}
\usepackage{amsmath}
\usepackage{hyperref}
\usepackage{amssymb}
\usepackage{graphicx}%
\begin{document}

\title{Scale Invariant Scattering in 2D}
\author{Thomas Curtright$^{\S }$ and Christophe Vignat$^{\circledcirc}$
\and {\footnotesize curtright@miami.edu
\ \ \ \ \ \ \ \ \ \ \ \ \ \ \ cvignat@tulane.edu \ \ \ \ \ \ }\medskip\\$^{\S }$Department of Physics, University of Miami, Coral Gables, FL 33124\\$^{\circledcirc}$Department of Mathematics, Tulane University, New Orleans, LA 70118}
\date{\vspace{-0.25in}}
\maketitle

\begin{abstract}
\textit{For a non-relativistic scale invariant system in two spatial
dimensions, the quantum scattering amplitude }$f\left(  \theta\right)
$\textit{ is given as a dispersion relation, with a simple closed form for
}$\operatorname{Im}f\left(  \theta\right)  $\textit{ as well as the integrated
cross-section }$\sigma\propto\operatorname{Im}f\left(  \theta=0\right)
$\textit{. \ For fixed }$\theta\neq0$, t\textit{he }$\hbar\rightarrow0$
\textit{classical} \textit{limit is straightforward to obtain.\bigskip}

\end{abstract}

It is well-known that non-relativistic scattering by a $V=\kappa/r^{2}$
potential is scale invariant \cite{Jackiw}, in any number of spatial
dimensions, and the scattering problem is well-defined mathematically
\cite{Newton} for all $\kappa>0$, even though the total integrated scattering
cross-section $\sigma$ is \emph{infinite} when computed classically. \ An
infinite $\sigma$ is also obtained for quantum mechanical scattering by a
$1/r^{2}$ potential in three spatial dimensions \cite{3DRemark}.

However, in two spatial dimensions the integrated cross-section, $\sigma
=\int_{0}^{2\pi}\left(  \frac{d\sigma}{d\theta}\right)  d\theta$, is
\emph{finite} when computed using quantum mechanics. \ The result for a
mono-energetic beam, with energy $E=\hbar^{2}k^{2}/\left(  2m\right)  $, is
\cite{TLC}
\begin{equation}
\sigma=\frac{\pi^{2}\kappa}{\hbar}\sqrt{\frac{2m}{E}} \label{SimpleSigma}%
\end{equation}
As an extension of this result, in this paper the exact scattering amplitude
$f\left(  \theta\right)  $ is discussed for a plane wave incident on the
potential. \ This amplitude is recast here as a dispersion relation, resulting
in a form amenable to analytic as well as numerical calculations. \ 

As given by phase shift analysis,
\begin{equation}
f\left(  \theta\right)  =\sqrt{\frac{2}{\pi k}}\sum_{l=-\infty}^{\infty
}e^{il\theta}e^{i\delta_{l}}\sin\left(  \delta_{l}\right)  \label{PSA}%
\end{equation}
where the phase shifts are exactly $\delta_{l}=\frac{\pi}{2}\left(
\sqrt{l^{2}}-\sqrt{l^{2}+2m\kappa/\hbar^{2}}\right)  $ with no energy
dependence, thereby exhibiting scale invariance in this context
\cite{CutoffRemark}. \ The differential cross-section is of course
$d\sigma/d\theta=\left\vert f\left(  \theta\right)  \right\vert ^{2}$ which
integrates to give the total cross-section%
\begin{equation}
\sigma=\frac{4}{k}\sum_{l=-\infty}^{\infty}\sin^{2}\left(  \delta_{l}\right)
=\frac{4}{k}\left(  \sin^{2}\left[  \frac{\pi}{2}\sqrt{2m\kappa/\hbar^{2}%
}\right]  +2\sum_{l=1}^{\infty}\sin^{2}\left[  \frac{\pi}{2}\left(
\sqrt{l^{2}+2m\kappa/\hbar^{2}}-l\right)  \right]  \right)  \label{Unsummed}%
\end{equation}
The final result (\ref{SimpleSigma}) for $\sigma$ therefore follows from the
evaluation of this last sum, as carried out in \cite{TLC}. \ 

Going beyond that, the sum in (\ref{PSA}) can be recast as the following
integral relation (i.e. dispersion relation a.k.a. Hilbert transform
\cite{King}) valid for $-\pi\leq\theta\leq\pi$.%
\begin{align}
\sqrt{\frac{2k}{\pi}}f\left(  \theta\right)   &  =\lim_{\varepsilon
\rightarrow0}\int_{\vartheta=0}^{\vartheta=\pi}\frac{xJ_{1}\left(
x\sqrt{\vartheta\left(  2\pi-\vartheta\right)  }\right)  }{\sqrt
{\vartheta\left(  2\pi-\vartheta\right)  }}\frac{\sin\vartheta}{\cos
\theta-\cos\vartheta-i\varepsilon}~d\vartheta\nonumber\\
&  \ \ \ \ \ +\int_{\tau=0}^{\tau=\infty}\frac{xJ_{1}\left(  x\sqrt{\pi
^{2}+\tau^{2}}\right)  }{\sqrt{\pi^{2}+\tau^{2}}}\frac{\sinh\tau}{\cos
\theta+\cosh\tau}~d\tau\label{DR2D}%
\end{align}
where $J_{1}$ is a Bessel function, in standard notation \cite{Numerics}.
\ The parameter $x$ is related to the potential strength.
\begin{equation}
x^{2}\equiv2m\kappa/\hbar^{2}>0\text{ \ \ if \ }V\left(  r\right)
=\kappa/r^{2}\text{ \ \ with \ \ }\kappa>0
\end{equation}
where $m$ is the mass of the particle.\newpage

\vspace*{0.5in}The first integral in (\ref{DR2D}) is a dispersion over
physically allowed momentum transfers:
\begin{equation}
\Delta k^{2}=\left(  \overrightarrow{k}_{in}-\overrightarrow{k}_{out}\right)
^{2}=2k^{2}\left(  1-\cos\theta\right)
\end{equation}
where $\theta$ is the scattering angle, and $\overrightarrow{k}_{in}%
^{2}=\overrightarrow{k}_{out}^{2}\equiv k^{2}$ . \ The second integral in
(\ref{DR2D}) is a dispersion over \emph{un}physical momentum transfers. \ The
result (\ref{DR2D}) is established by some straightforward analysis
\cite{TCCV}. \ 

Given the result, the total integrated cross-section is obtained from the
imaginary part of $f\left(  \theta\right)  $ through the 2D optical theorem.
\ From the operational calculus identity $\lim_{\varepsilon\rightarrow
0}1/\left(  \cos\theta-\cos\vartheta-i\varepsilon\right)  =P\left(  \frac
{1}{\cos\theta-\cos\vartheta}\right)  +i\pi\delta\left(  \cos\vartheta
-\cos\theta\right)  $, where $P$ means Cauchy principal value integration and
$\delta$ is a Dirac delta, it follows that the imaginary part of the amplitude
is given entirely by the Dirac delta contribution.
\begin{equation}
\sqrt{\frac{\pi k}{2}}\operatorname{Im}f\left(  \theta\right)  =\frac{\pi
^{2}xJ_{1}\left(  x\sqrt{\left\vert \theta\right\vert \left(  2\pi-\left\vert
\theta\right\vert \right)  }\right)  }{2\sqrt{\left\vert \theta\right\vert
\left(  2\pi-\left\vert \theta\right\vert \right)  }} \label{ImAmp}%
\end{equation}
Therefore, from $\sigma=\int_{-\pi}^{\pi}\left\vert f\left(  \theta\right)
\right\vert ^{2}d\theta=\frac{4}{k}\sum_{l=-\infty}^{\infty}\sin^{2}\left(
\delta_{l}\right)  =\frac{4}{k}\sqrt{\frac{\pi k}{2}}\operatorname{Im}f\left(
\theta=0\right)  $, as well as the small argument behavior $J_{1}\left(
z\right)  =\frac{1}{2}z+O\left(  z^{3}\right)  $, the 2D integrated
cross-section is given exactly by
\begin{equation}
\sigma=\frac{4}{k}\sqrt{\frac{\pi k}{2}}\operatorname{Im}f\left(
\theta=0\right)  =\frac{\pi^{2}}{k}\frac{2m\kappa}{\hbar^{2}}%
\end{equation}
in agreement with (\ref{SimpleSigma}).

Moreover, the classical differential cross-section is evident as the
$\hbar\rightarrow0$ limit of the QM differential cross-section
\cite{StrongCouplingRemark}. \ In that limit, $x=\sqrt{\frac{2m\kappa}%
{\hbar^{2}}}\underset{\hbar\rightarrow0}{\longrightarrow}\infty$, so from
(\ref{ImAmp}) and the asymptotic behavior $J_{1}\left(  z\right)
\underset{z\rightarrow\infty}{\sim}\sqrt{\frac{2}{\pi z}}\cos\left(
z-3\pi/4\right)  $, for fixed $\theta\neq0$,
\begin{equation}
\operatorname{Im}f\left(  \theta\right)  \underset{\hbar\rightarrow0}{\sim
}\left(  \frac{2m\kappa}{\hbar^{2}k^{2}}\right)  ^{1/4}~~\frac{\pi\cos\left(
x\sqrt{\left\vert \theta\right\vert \left(  2\pi-\left\vert \theta\right\vert
\right)  }-3\pi/4\right)  }{\left(  \left\vert \theta\right\vert \left(
2\pi-\left\vert \theta\right\vert \right)  \right)  ^{3/4}}\label{Imf}%
\end{equation}
In this same limit it takes a bit of coaxing \cite{TCCV} to obtain the
asymptotic form of $\operatorname{Re}f\left(  \theta\right)  $ from the
dispersion relation (\ref{DR2D}), but in view of $\operatorname{Im}f\left(
\theta\right)  $ the result is hardly surprising \cite{DoItYourSelf}.%
\begin{equation}
\operatorname{Re}f\left(  \theta\right)  \underset{\hbar\rightarrow0}{\sim
}-\left(  \frac{2m\kappa}{\hbar^{2}k^{2}}\right)  ^{1/4}~~\frac{\pi\sin\left(
x\sqrt{\left\vert \theta\right\vert \left(  2\pi-\left\vert \theta\right\vert
\right)  }-3\pi/4\right)  }{\left(  \left\vert \theta\right\vert \left(
2\pi-\left\vert \theta\right\vert \right)  \right)  ^{3/4}}\label{Ref}%
\end{equation}
Hence the differential cross-section in the classical limit is
\begin{equation}
\frac{d\sigma}{d\theta}=\left(  \operatorname{Re}f\left(  \theta\right)
\right)  ^{2}+\left(  \operatorname{Im}f\left(  \theta\right)  \right)
^{2}\underset{\hbar\rightarrow0}{\sim}\left(  \frac{\kappa}{E}\right)
^{1/2}~\frac{\pi^{2}}{\left(  \left\vert \theta\right\vert \left(
2\pi-\left\vert \theta\right\vert \right)  \right)  ^{3/2}}%
\end{equation}
so long as $\theta\neq0$, where the classical momentum is $\hbar k=\sqrt{2mE}%
$. \ This agrees exactly with the 2D differential cross-section obtained from
classical mechanics.

The classical $d\sigma/d\theta$ diverges as $\theta^{-3/2}$ for forward
scattering, but the quantum $d\sigma/d\theta$ diverges only as $\ln^{2}\theta
$, where the latter behavior follows from the $\ln\theta$ singularity in
$\operatorname{Re}f\left(  \theta\right)  $ as given by the first integral in
(\ref{DR2D}). \ Thus the integrated $\sigma$ is finite for the quantum system,
but not for the classical system. \ This is also evident from the
$\hbar\rightarrow0$\ limit of (\ref{SimpleSigma}).\bigskip

\textbf{Acknowledgements} \ We thank T.S. Van Kortryk and G. Verma for
checking some of the math. \ TC received financial support from the United
States Social Security Administration.


\begin{thebibliography}{99}
\bibitem {3DRemark}This follows from partial wave analysis similar to the 2D
case, as given in the following, except in 3D the phase shift is $\delta
_{l}=\frac{\pi}{2}\left(  \sqrt{l\left(  l+1\right)  }-\sqrt{l\left(
l+1\right)  +2m\kappa/\hbar^{2}}\right)  $ and the sum $\sum_{l=0}^{\infty
}\left(  2l+1\right)  \sin^{2}\delta_{l}$ is then divergent because of the
additional $2l+1$ multiplicity factor.
\bibitem {CutoffRemark}If the potential is cut-off outside some large finite
radius, $R$, such that $V=\kappa\Theta\left(  R-r\right)  /r^{2}$ where
$\Theta$ is the Heaviside step function, then the scattering amplitudes are
given in terms of Bessel functions as a ratio of Wronskians,%
\[
e^{2i\delta_{l}}=-W\left[  J_{\nu\left(  l\right)  }\left(  kR\right)
,H_{l}^{\left(  2\right)  }\left(  kR\right)  \right]  /W\left[  J_{\nu\left(
l\right)  }\left(  kR\right)  ,H_{l}^{\left(  1\right)  }\left(  kR\right)
\right]
\]
where $\nu\left(  l\right)  =\sqrt{l^{2}+2m\kappa/\hbar^{2}}$, with the result
$\delta_{l}=\frac{\pi}{2}\left(  \sqrt{l^{2}}-\sqrt{l^{2}+2m\kappa/\hbar^{2}%
}\right)  $ recovered in the limit $kR\rightarrow\infty$.
\bibitem {Numerics}A small angle singularity (i.e. $\ln\theta$) in
$\operatorname{Re}f\left(  \theta\right)  $ is manifest upon writing the
Cauchy principal value integral (i.e. $C\hspace{-0.13in}%
{\displaystyle\int}
$) implicit in (\ref{DR2D})\ as a once-subtracted dispersion relation:
\begin{multline*}
C\hspace{-0.15in}\int_{\vartheta=0}^{\vartheta=\pi}\frac{xJ_{1}\left(
x\sqrt{\vartheta\left(  2\pi-\vartheta\right)  }\right)  }{\sqrt
{\vartheta\left(  2\pi-\vartheta\right)  }}\frac{\sin\vartheta}{\cos
\theta-\cos\vartheta}~d\vartheta=\frac{xJ_{1}\left(  x\sqrt{\theta\left(
2\pi-\theta\right)  }\right)  }{\sqrt{\theta\left(  2\pi-\theta\right)  }}%
~\ln\left(  \frac{1+\cos\theta}{1-\cos\theta}\right) \\
+\int_{\vartheta=0}^{\vartheta=\pi}\left(  \frac{xJ_{1}\left(  x\sqrt
{\vartheta\left(  2\pi-\vartheta\right)  }\right)  }{\sqrt{\vartheta\left(
2\pi-\vartheta\right)  }}-\frac{xJ_{1}\left(  x\sqrt{\theta\left(  2\pi
-\theta\right)  }\right)  }{\sqrt{\theta\left(  2\pi-\theta\right)  }}\right)
\frac{\sin\vartheta}{\cos\theta-\cos\vartheta}~d\vartheta
\end{multline*}
where the final integration is finite for $0\leq\theta\leq\pi$, and is
well-suited for numerical evaluation. \ Also note that the singularity
exhibited by this result at $\theta=\pi$ is eliminated upon combining both
integrals in (\ref{DR2D}). \ The cancellation in question is manifest upon
also expressing the second integral in (\ref{DR2D}) as a once-subtracted
dispersion relation:
\begin{gather*}
\int_{\tau=0}^{\tau=\infty}\frac{xJ_{1}\left(  x\sqrt{\pi^{2}+\tau^{2}%
}\right)  }{\sqrt{\pi^{2}+\tau^{2}}}\frac{\sinh\tau}{\cosh\tau+\cos\theta
}~d\tau=\frac{1}{\pi}\sin\left(  \pi x\right)  -\int_{\tau=0}^{\tau=\infty
}\frac{xJ_{1}\left(  x\sqrt{\pi^{2}+\tau^{2}}\right)  }{\sqrt{\pi^{2}+\tau
^{2}}}\frac{e^{-\tau}+\cos\theta}{\cosh\tau+\cos\theta}~d\tau\\
=\frac{1}{\pi}\sin\left(  \pi x\right)  -\frac{xJ_{1}\left(  x\pi\right)
}{\pi}\left(  \ln\left(  1+\cos\theta\right)  +\ln2\right)  +\int_{\tau
=0}^{\tau=\infty}\left(  \frac{xJ_{1}\left(  x\pi\right)  }{\pi}-\frac
{xJ_{1}\left(  x\sqrt{\pi^{2}+\tau^{2}}\right)  }{\sqrt{\pi^{2}+\tau^{2}}%
}\right)  \frac{e^{-\tau}+\cos\theta}{\cosh\tau+\cos\theta}~d\tau
\end{gather*}
Once again, the last integration is finite for $0\leq\theta\leq\pi$ and
well-suited for numerical evaluation.
\bibitem {StrongCouplingRemark}The same result is obtained for fixed $\hbar
$\ in the strong coupling limit $\kappa\rightarrow\infty$.
\bibitem {DoItYourSelf}For a conscientious reader, who wishes to verify the
asymptotics of $\operatorname{Re}f\left(  \theta\right)  $ starting from the
Hilbert transform (\ref{DR2D}), we suggest consulting Chapters 8 \& 11\ in
\cite{King}, Volume 1.
\end{thebibliography}
\end{document}